# Huygens-Fresnel principle for molecular continuum wave function

A. S. Baltenkov

Arifov Institute of Electronics, 100125, Tashkent, Uzbekistan

**Abstract**

The asymptotic behavior of the molecular continuum wave function has been analyzed within a model of non-overlapping atomic potentials. It has been shown that the representation of the wave function far from a molecule as a plane wave and single spherical wave emitted by the molecular center cannot be corrected. Because of the multicenter character of the problem, the asymptotic form of the wave function according to the Huygens-Fresnel principle must contain $N$ spherical waves with centers at the nuclei of the $N$ atoms that form the molecule. A method of partial waves for a spherically non-symmetrical target is considered for the simplest multicenter target formed by two non-overlapping potentials. The results are compared with those obtained within the single spherical wave approximation. It has been shown that the use of this approximation is intrinsically conflicting, which is a direct consequence of refusal from the Huygens-Fresnel picture of the wave scattering process.

## 1. Introduction

The multiple scattering (MS) methodology is one of the most popular theoretical constructions used for calculation of molecular continuum wave functions. The general ideas of this method were developed by Dill and Dehmer in paper [1] where "the multiple scattering technique for treating nonseparable eigenvalue problems with electron-scattering theory to construct continuum wave functions" was combined. The MS method [1] and its later modifications [2-7] are widely used now to calculate the cross sections of electron elastic scattering by molecules and those of molecular photoionization. Originally the MS method was used in molecular physics to calculate bound state eigenvalues [8]. For the calculation of the bound state wave functions their normalization is evident. The situation with the continuum wave functions is quite different. A choice of their normalization, i.e. the asymptotic behavior of the wave functions, requires to be analyzed. This moment is of great importance for the accuracy of any method of molecular continuum calculations to be estimated, particularly when one deals with differential cross sections of fixed-in-space targets, because these cross sections are extremely sensitive to the asymptotic behavior of the wave function.

One of the general ideas of the methods [1-7] is an assumption that the asymptotic form of the wave function far from the molecule is a sum of a plane wave plus a single spherical wave (SSW) emitted by the molecular center and therefore the radial parts of the electron wave functions outside so-called "molecular sphere" can be represented as a linear combination of the regular and irregular solutions of the Schrödinger equation for the potential that in this region "is taken to be spherical … about the molecular center" [1]. The coefficients of this linear combination are defined by the molecular phases of scattering. They are defined by the matching conditions of the continuum wave function on the surfaces of the atomic and molecular spheres. Hence, in the method [1] the solution of the problem of electron scattering by a spherically non-symmetrical potential is reduced to the usual method of the partial waves for a spherical target. Proposed in paper [1], the recipes to build a continuum wave function outside the molecular sphere are considered as a matter-of-course and, as far as we know, they are beyond any doubt. The picture of the wave scattering in the base of the MS method is given in Fig. *1a*. In this muffin-tin potential picture the bound electrons are located inside the atomic spheres (Regions $I_1$ and $I_2$). The latter are confined to an outer molecular sphere that surrounds the molecule as a whole. Beyond the molecular sphere (Region III) there is a single spherical wave emitted by the molecular center.

Similar with Fig. 1*a*, the scattering picture is in the base of another method of molecular continuum calculation – so-called single-center-expansion (SCE) approach [9-14]. In this method the interaction potentials of electrons with atomic nuclei and the wave functions are expanded in a series in spherical functions around the single center, normally the center of mass of the system. As a result of such expansion, the space is divided into two parts: internal where the electron wave function is described by the zero-regular solution of the wave equation and external one where this function is a linear combination of the regular and irregular solutions of the radial Schrödinger equation. Matching of these functions defines the coefficients of this linear combination and molecular phases of scattering. Such a representation of the interaction potentials and wave functions makes matrix element calculations much easier. As in the MS method, the SCE approach is originally used to calculate the bound states of electrons in molecules [15]. As far as the molecular continuum is concerned, the SCE-method similar to the MS methodology reduces the problem of electron scattering by a spherically non-symmetrical molecular potential to the usual method of the partial waves for a spherical target. In [11] (page 383) we read: "the SCE approach makes the numerical procedure analogous to e-atom (ion) scattering". For this reason, main attention in both methods is paid to calculating the radial parts of the wave functions from which the scattering wave function is constructed by analogy with the spherically symmetrical case.

It is evident that such an approach to describing the process of electron wave scattering by a multicenter target contradicts the classical physical picture of wave scattering [16-22] which is based on the Huygens-Fresnel principle. According to this principle, in scattering of the initial wave by the system of scatterers each of them becomes a source of the secondary spherically scattered waves and far from the target there is a system of the spherical waves diverging from each of the centers as shown in Fig. 1 *b*. It is known that the interference of the spherical waves creates a diffraction pattern whose properties depend periodically on the ratio of the inter-nuclear distance to the electron wavelength. If we suppose that far from the system of the scattering centers there is a single spherical wave then the phenomenon of electron diffraction by molecules as the interference of a few spherical waves emitted by the spatially separated sources becomes impossible at all. Therefore it is difficult to expect that SSW approach (Fig. 1*a*) can be the basis for correct description of molecular continuum wave functions.

In the connection with the above stated some questions arise: What are the consequences of refusal from the Huygens-Fresnel principle while calculating the molecular continuum wave functions? Is it possible or not to adapt the method of partial waves for the case of a multicenter target formed by non-overlapping atomic potentials keeping the Huygens-Fresnel picture of the scattering process? The present paper is devoted to analysis of these questions. Here for a simple example of slow particle scattering by short-range non-overlapping potentials we will analyze the special features of the partial wave method for non-spherical targets and compare the results obtained with those following from the SSW asymptotic. In Sec. 2 we will consider slow particle scattering by central potential shifted from the coordinate origin. Consideration of scattering for the simplest spherically non-symmetrical target makes it possible to reveal the main difficulties of the SSW approach. In Sec. 3 the similar analysis will be performed for a target formed by two short-range potentials. In Sec. 4 the S-matrix method for non-spherical potentials developed in [23] will be applied for this two-center target. Sec. 5 is conclusions.

## 2. One-center target

It is known that the wave function describing elastic scattering of a particle (electron) by a spherically symmetrical potential is defined by the expression [24]

$$\psi_{\mathbf{k}}^{+}(\mathbf{r}) = 4\pi \sum_{l=0}^{\infty} R_{kl}(r) Y_{lm}^{*}(\mathbf{k}) Y_{lm}(\mathbf{r}),  \qquad (1)$$

where the radial part of the wave function has the asymptotic form

$$R_{kl}(r \to \infty) \approx i^{l} e^{i\eta_{l}} \frac{1}{kr} \sin(kr - \frac{\pi l}{2} + \eta_{l}). \qquad (2)$$

Such representation of the wave function separates in the explicit form the scattering dynamics contained in the molecular phases $\eta_\lambda(k)$ from the kinematics of the process defined by the spherical functions $Y_{ml}(\mathbf{k})$ and $Y_{ml}(\mathbf{r}) \equiv Y_{ml}(\mathbf{k'})$. Here $Y_{lm}(\mathbf{r}) \equiv Y_{lm}(\vartheta, \varphi)$ where $\vartheta$ and $\varphi$ are the spherical angles of the vector $\mathbf{r}$. In the case of a target formed by several non-overlapping atomic potentials the scattering wave function much include, besides the electron wave vectors before and after collision: $\mathbf{k}$ and $\mathbf{k'} = k\mathbf{r}/r$, a set of the vectors $\{\mathbf{R}\}$ defining the position of these potentials in space. If one supposes, as it is for the MS and SCE methods, that the radial part of the wave function can be represented as a linear combination of the regular and irregular solutions, i.e. its asymptotic is defined as before by the formula (2) then the scattering phases $\eta_\lambda(k)$ must be the functions of these vectors as well. Let us show that this is impossible.

It is known that the exact wave function defining the *s*-scattering of slow particle (electron) by short-range potential being at the coordinate origin beyond the range of potential action has the form [24,25]

$$\psi_\mathbf{k}^+(\mathbf{r}) = e^{i\mathbf{k}\cdot\mathbf{r}} + CG_k^+(\mathbf{r},0) = e^{i\mathbf{k}\cdot\mathbf{r}} + \frac{C}{2\pi}\frac{e^{ikr}}{r}, \qquad r > \rho. \tag{3}$$

Here $\rho$ is the radius of potential action, $G_k^+(\mathbf{r},\mathbf{r'})$ is the free particle Green function obeyed the equations

$$-\frac{1}{2}(\nabla^2 + k^2)G_k^+(\mathbf{r},\mathbf{r'}) = \delta(\mathbf{r}-\mathbf{r'}), \qquad G_k^+(\mathbf{r},\mathbf{r'}) = \frac{e^{ik|\mathbf{r}-\mathbf{r'}|}}{2\pi|\mathbf{r}-\mathbf{r'}|}. \tag{4}$$

Throughout the paper we use the atomic system of units. The wave function (3) is a superposition of the plane wave incident onto the center and the diverging spherical wave of *s*-type $G_k^+(\mathbf{r},0) = \frac{1}{\sqrt{2\pi}}\frac{e^{ikr}}{r}Y_{00}(\mathbf{r})$ with the center at the coordinate origin. The coefficient $C$ in formula (3) can be defined either from the boundary condition imposed on the wave function at the coordinate origin

$$\frac{1}{r\psi_\mathbf{k}^+(\mathbf{r})}\frac{d[r\psi_\mathbf{k}^+(\mathbf{r})]}{dr}\bigg|_{r\to 0} = k\cot\delta_0, \tag{5}$$

or by way of introducing a pseudopotential in the wave equation [25,26]

$$(\nabla^2 + k^2)\psi_\mathbf{k}^+(\mathbf{r}) = -\frac{4\pi}{k\cot\delta_0}\delta(\mathbf{r})\frac{\partial}{\partial r}[r\psi_\mathbf{k}^+(\mathbf{r})]. \tag{6}$$

In these formulas $\delta_0$ is the *s*-phase of scattering by target potential. Substituting the function (3) to the equations (5) and (6), we obtain for the coefficient $C$ the following formula

$$C = \frac{2\pi}{k(\cot\delta_0 - i)}. \tag{7}$$

From (7) for the scattering amplitude $F(\mathbf{k},\mathbf{k'})$ and the total scattering cross section $\sigma(k)$ we have the following known expressions

$$F(\mathbf{k},\mathbf{k'}) = \frac{1}{k(\cot\delta_0 - i)} = \frac{1}{k}e^{i\delta_0}\sin\delta_0, \qquad \sigma(k) = \frac{4\pi}{k^2}\sin^2\delta_0. \tag{8}$$

Let us write the asymptotic of the wave function (3) as expansion in a partial series in spherical functions with the center at the coordinate origin and define the relation between the atomic phase $\delta_0$ and the infinite set of "molecular" phases $\eta_l$ determining the behavior of the wave function for $r \to \infty$

$$e^{i\mathbf{k}\cdot\mathbf{r}} + F(\mathbf{k},\mathbf{k'})\frac{e^{ikr}}{r} = 4\pi\sum_{lm}i^l e^{i\eta_l}\frac{1}{kr}\sin(kr - \frac{\pi l}{2} + \eta_l)Y_{lm}(\mathbf{r})Y_{lm}^*(\mathbf{k}) \tag{9}$$

Expressing the exponent in (9) as a series in spherical functions and using by the usual way their orthogonality, we obtain the following equality

$$i^l[\sin(kr - \frac{\pi l}{2}) - e^{i\eta_l}\sin(kr - \frac{\pi l}{2} + \eta_l)] = -kF(\mathbf{k},\mathbf{k}')e^{ikr}\delta_{l0}\delta_{m0}. \tag{10}$$

Here $\delta_{l0}$ and $\delta_{m0}$ are the Kroneker symbols. From the equation (10) we obtain for the phases $\eta_l$, as it should be expected, the following values

$$\eta_l = \begin{cases} \delta_0 & \text{for } l = 0, \\ 0 & \text{for } l > 0. \end{cases} \tag{11}$$

Now let us consider the case of the same potential but shifted from the coordinate origin to the point **R**. The scattering wave function, according to the Huygens-Fresnel principle, will be found as a plane wave and spherical *s*-wave with the center at the point **R**:

$$\psi_{\mathbf{k}}^+(\mathbf{r}) = e^{i\mathbf{k}\cdot\mathbf{r}} + CG_k^+(\mathbf{r},\mathbf{R}). \tag{12}$$

Substituting this function to the wave equation

$$(\nabla^2 + k^2)\psi_{\mathbf{k}}^+(\mathbf{r}) = -\frac{4\pi}{k\cot\delta_0}\delta(\mathbf{r}-\mathbf{R})\frac{\partial}{\partial\rho}[\rho\psi_k^+(\mathbf{r})], \qquad \rho = |\mathbf{r}-\mathbf{R}|, \tag{13}$$

we obtain the following expression for the coefficient *C*

$$C = \frac{2\pi e^{i\mathbf{k}\cdot\mathbf{R}}}{k(\cot\delta_0 - i)}. \tag{14}$$

From (14) for the amplitude of scattering by the shifted potential we obtain the expression

$$F(\mathbf{k},\mathbf{k}',\mathbf{R}) = \frac{e^{i(\mathbf{k}-\mathbf{k}')\cdot\mathbf{R}}}{k(\cot\delta_0 - i)} = F(\mathbf{k},\mathbf{k}')e^{i(\mathbf{k}-\mathbf{k}')\cdot\mathbf{R}}, \tag{15}$$

The amplitude (15) differs from (8) by the factor $\exp[i(\mathbf{k}-\mathbf{k}')\cdot\mathbf{R}]$ the presence of which has no effect on the scattering characteristics defined by the square of scattering amplitude modulo. The differential and total cross sections of scattering by the potential with the center at **R**, as it should be, coincide with those of scattering by a target at the coordinate origin.

The partial expansion of the wave function asymptotic (12) as a series in spherical functions with the center at the coordinate origin results in the following relation

$$e^{i\mathbf{k}\cdot\mathbf{r}} + F(\mathbf{k},\mathbf{k}')e^{i(\mathbf{k}-\mathbf{k}')\cdot\mathbf{R}}\frac{e^{ikr}}{r} = 4\pi\sum_{lm}i^l e^{i\eta_l}\frac{1}{kr}\sin(kr - \frac{\pi l}{2} + \eta_l)Y_{lm}(\mathbf{r})Y_{lm}^*(\mathbf{k}). \tag{16}$$

From here after integration over angles of the vectors **r** and **k** we obtain the following equation connecting $\delta_0$ with the phases $\eta_l$:

$$i^l[\sin(kr - \frac{\pi l}{2}) - e^{i\eta_l}\sin(kr - \frac{\pi l}{2} + \eta_l)] = -4\pi kF(\mathbf{k},\mathbf{k}')e^{ikr}j_l^2(kR)|Y_{lm}(\mathbf{R})|^2. \tag{17}$$

Here $j_l(kR)$ is the spherical Bessel function. From the equation (17) it follows that the phases $\eta_l$ and $\delta_0$ are connected by the equation

$$e^{i\eta_l}\sin\eta_l = 4\pi j_l^2(kR)|Y_{lm}(\mathbf{R})|^2 e^{i\delta_0}\sin\delta_0. \tag{18}$$

For the shift $R = 0$ the only non-zero Bessel function is $j_0(kR)$. In this case the relation of the phases $\delta_0$ and $\eta_l$ is defined by the formula (11). For the non-zero shift ($\mathbf{R} \neq 0$) the Bessel functions in (18) differ from zero. Equating the real and imaginary parts in the left and right sides of the equation (18), we have the following relation between the phases

$$\eta_l = \delta_0, \qquad \text{for all } l. \tag{19}$$

The phases $\eta_l$ are independent of the orbital moment *l* and their number is infinite, therefore the total cross section of scattering by shifted center is equal to infinity

$$\sigma(k) = \frac{4\pi}{k^2}\sum_{l=0}^{\infty}(2l+1)\sin^2\eta_l = \frac{4\pi}{k^2}\sin^2\delta_0\sum_{l=0}^{\infty}(2l+1) = \infty. \tag{20}$$

Hence, the attempt to represent the wave function of scattering by center shifted from the coordinate origin as a series in spherical functions with the center at the coordinate origin leads to unavoidable principal contradictions and for this reason the partial expansion (16) and representation of the radial part of wave function as a linear combination of the regular and irregular solutions of the wave equation are impossible.

It is known that in the centrally symmetrical case the elastic scattering amplitude is defined by two vectors $\mathbf{k}$ and $\mathbf{k'}$. The amplitude dependence on these vectors is in the spherical functions $Y_{lm}(\mathbf{k})$ and $Y_{lm}(\mathbf{k'})$ while the scattering phase depends on modulo of these vectors only. While shifting the potential from the coordinate origin an additional vector $\mathbf{R}$ enters the play. If one supposes that the representation of scattering function as a series in spherical functions with the center at the coordinate origin is correct then the dependence on vector $\mathbf{R}$ can be only in the scattering phases as it follows from (16). However, there are no real solutions of this equation for $\eta_l(\mathbf{R})$. Therefore, the transfer of the wave function dependence on vector $\mathbf{R}$ to the molecular scattering phases $\eta_l$ is impossible.

In connection with the above stated a question arises whether this conclusion is a consequence of the special features of the above considered potential. Let us show that this is not so. The amplitude of elastic scattering by *any* short-range spherically symmetrical potential $V(r)$ in the Born approximation is defined by the expression [24]

$$F^B(\mathbf{k},\mathbf{k'}) = -\frac{1}{2\pi}\int e^{i(\mathbf{k}-\mathbf{k'})r}V(r)d\mathbf{r}. \qquad (21)$$

For the same potential shifted from the coordinate origin to the point $\mathbf{R}$ by replacing integration variable we obtain for the scattering amplitude the following expression

$$F^B(\mathbf{k},\mathbf{k'},\mathbf{R}) = -\frac{1}{2\pi}\int e^{i(\mathbf{k}-\mathbf{k'})\cdot\mathbf{r}}V(\mathbf{r}-\mathbf{R})d\mathbf{r} = F^B(\mathbf{k},\mathbf{k'})e^{i(\mathbf{k}-\mathbf{k'})\cdot\mathbf{R}} \qquad (22)$$

exactly coinciding with (15). Let us write the left side of this equation as usual amplitude of scattering by spherically symmetrical target. Now instead of (22) we obtain the following equation

$$\frac{2\pi}{ik}\sum_{lm}(e^{2i\eta_l}-1)Y_{lm}(\mathbf{k})Y^*_{lm}(\mathbf{k'})e^{i\mathbf{k'}\cdot\mathbf{R}} = F^B(\mathbf{k},\mathbf{k'})e^{i\mathbf{k}\cdot\mathbf{R}}. \qquad (23)$$

Writing the amplitude $F^B(\mathbf{k},\mathbf{k'})$ as a series in spherical functions

$$F^B(\mathbf{k},\mathbf{k'}) = \sum_{\lambda\mu} F_{\lambda\mu}(k)Y_{\lambda\mu}(\mathbf{k})Y^*_{\lambda\mu}(\mathbf{k'}), \qquad (24)$$

integrating the both sides of equation (23) first over all the angles of the vector $\mathbf{k'}$ and then the vector $\mathbf{k}$, we obtain the following equation for molecular phases $\eta_l$

$$\frac{2\pi}{ik}(e^{2i\eta_l}-1) = F_{00}(k). \qquad (25)$$

According to (25), for fixed momentum of scattering particle $k$ the phases are constant and independent of orbital moment. Consequently, we again come to the conclusion that the cross section of scattering by shifted potential is infinite; i.e. for arbitrary shape of the potential $V(r)$ the transfer of the dependence on vector $\mathbf{R}$ into scattering phases is impossible.

The MS and SCE methods operate with multicenter targets whose scattering amplitude (in the first approximation) is the sum of the scattering amplitudes of the single centers, each of these amplitudes being multiplied by a translation factor $\exp(i\mathbf{k}\cdot\mathbf{R}_N)$ taking into account the phase differences of incident wave due to the spatial disposition of the $N$ scatterers. Therefore, the molecular phases $\eta_l$ in these methods should be functions of a set of $N$ vectors $\mathbf{R}_N$ rather than one vector $\mathbf{R}$. Since this is impossible even for one vector, it becomes evident that for the scattering phases, as in the centrally symmetrical case, to be functions of particle momentum $k$ only, it is necessary to transfer the dependencies on vectors $\mathbf{R}_N$ into some functions $Z(\mathbf{R}_N)$. The scattering wave function $\psi^+_\mathbf{k}(\mathbf{r})$ should be developed as a series in these functions. They, similar to spherical functions, should form a complete set of orthonormal functions and their form will be defined by a specific structure of the multi-atomic target. We will come back to these functions in Sec. 4.

## 3. Two-center problem

Consider the scattering of a slow electron by two identical non-overlapping atomic potentials with the centers at $\mathbf{r} = \pm \mathbf{R}/2$. This simplest multicenter system is a good example illustrating contradictions appearing in attempt to apply the usual partial waves method for non-spherical targets formed by non-overlapping atomic potentials. It is important that the problem of slow particle scattering by this system of potentials can be solved analytically and so it can serve as a touchstone to analyze correctness of different calculation methods.

Beyond the action of the short-range potentials the wave function of particle scattered by this system has the form (see for example [21] where this function was used to describe scattering of slow mesons by deuterons)

$$\psi_{\mathbf{k}}^{+}(\mathbf{r}) = e^{i\mathbf{k}\cdot\mathbf{r}} + C_1(\mathbf{k})\frac{e^{ik|\mathbf{r}+\mathbf{R}/2|}}{|\mathbf{r}+\mathbf{R}/2|} + C_2(\mathbf{k})\frac{e^{ik|\mathbf{r}-\mathbf{R}/2|}}{|\mathbf{r}-\mathbf{R}/2|}, \qquad |\mathbf{r}\pm\mathbf{R}/2| > \rho. \qquad (26)$$

Here $\rho$ as before is the short-range potential radius and the coefficients at the spherical waves have the form [21,27]

$$C_1(\mathbf{k}) = \frac{ad - bd^*}{a^2 - b^2}; \qquad C_2(\mathbf{k}) = \frac{ad^* - bd}{a^2 - b^2};$$

$$a = e^{ikR}/R; \quad d = -e^{i\mathbf{k}\cdot\mathbf{R}/2}; \quad b = ik - k\cot\delta_0 = ik + q. \qquad (27)$$

The phase $\delta_0(k)$ is the $s$-wave phase for scattering by each of the atomic potentials forming the target. The wave function (26) is the general solution of the Schrödinger equation and describes multiple scattering of a particle by two identical potentials. This function has the form of a superposition of plane wave plus *two* spherical $s$-waves generated by each of the target atoms. The function (26) corresponds to the Huygens-Fresnel pattern of scattering according to which the wave scattering by a system of the $N$ centers is accompanied by generation of $N$ secondary spherical waves.

The amplitude of the slow particle scattering by the target is obtained by considering the asymptotic behavior of the wave function (26)

$$F(\mathbf{k},\mathbf{k}',\mathbf{R}) = \frac{2}{a^2-b^2}\{b\cos[(\mathbf{k}-\mathbf{k}')\cdot\frac{\mathbf{R}}{2}] - a\cos[(\mathbf{k}+\mathbf{k}')\cdot\frac{\mathbf{R}}{2}]\}. \qquad (28)$$

Let us investigate the behavior of the amplitude (28) for great distances between centers:

$$F(\mathbf{k},\mathbf{k}',\mathbf{R}\to\infty) = F(\mathbf{k},\mathbf{k}')[e^{i(\mathbf{k}-\mathbf{k}')\cdot\mathbf{R}/2} + e^{-i(\mathbf{k}-\mathbf{k}')\cdot\mathbf{R}/2}]. \qquad (29)$$

According to (29), the amplitude of scattering by two space-apart centers is a sum of the amplitudes for single potentials with translation factors due to different special position of the scattering centers. The differences between formulas (28) and (29) are connected with the factor that (28) takes into account the processes of multiple scattering of particle by the centers for finite distances $R$ between them.

The scattering cross section is obtained from the amplitude (28) with the help of the optical theorem [24]

$$\sigma(k,\mathbf{R}) = \frac{4\pi}{k}\mathrm{Im}F(\mathbf{k}=\mathbf{k}',\mathbf{R}) = \frac{8\pi}{k}\mathrm{Im}\left[\frac{b-a\cos(\mathbf{k}\cdot\mathbf{R})}{a^2-b^2}\right]. \qquad (30)$$

We introduce the vector $\mathbf{R}$ in the argument of the cross section (30) to underline that we deal with the fixed-in-space molecule. The total cross section averaged over all the directions of momentum of incident electron $\mathbf{k}$ has the form:

$$\bar{\sigma}(k) = \frac{1}{4\pi}\int\sigma(k,\mathbf{R})d\Omega_k = \frac{8\pi}{k}\mathrm{Im}\left[\frac{b-aj_0(kR)}{a^2-b^2}\right]. \qquad (31)$$

Here $j_0(x) = \sin x/x$ is the spherical Bessel function.

Using the explicit expressions (27) for the functions $a$ and $b$, we obtain the following formula for the averaged cross section

$$\bar{\sigma}(k) = \frac{4\pi}{k^2}\left\{\left[1+\left(\frac{qR+\cos kR}{kR+\sin kR}\right)^2\right]^{-1} + \left[1+\left(\frac{qR-\cos kR}{kR-\sin kR}\right)^2\right]^{-1}\right\}. \qquad (32)$$

Let us investigate now what are consequences we will have if one supposes (as it is done in the MS and SCE methods) that the wave function (24) far from the target has the form SSW with the center at the coordinate origin. For this as before we write the asymptotic of the wave function (24) with the amplitude (26) as the partial expansion (1)

$$e^{i\mathbf{k}\cdot\mathbf{r}} + F(\mathbf{k},\mathbf{k}'\mathbf{R})\frac{e^{ikr}}{r} = 4\pi \sum_{lm} i^l e^{i\eta_l} \frac{1}{kr}\sin(kr - \frac{\pi l}{2} + \eta_l) Y_{lm}(\mathbf{r}) Y_{lm}^*(\mathbf{k}). \qquad (33)$$

Define from this equation the molecular phases of scattering. Integrating over vectors $\mathbf{k}$ and $\mathbf{k}'$ the both parts (33) and taking into account that

$$\iint F(\mathbf{k},\mathbf{k}',\mathbf{R}) Y_{lm}^*(\mathbf{k}') Y_{lm}(\mathbf{k}) = 32\pi^2 j_l^2(kR/2) |Y_{lm}(\mathbf{R})|^2 \frac{b-(-1)^l a}{a^2 - b^2}, \qquad (34)$$

we obtain the following equation for molecular phases $\eta_l(k)$:

$$e^{i\eta_l}\sin\eta_l = 8\pi k j_l^2(kR/2) |Y_{lm}(\mathbf{R})|^2 \frac{b-(-1)^l a}{a^2 - b^2} = A_l \frac{b-(-1)^l a}{a^2 - b^2}. \qquad (35)$$

From equality of the real and imaginary parts of the equation (35) we obtain for molecular phases the following expressions

$$\cot\eta_l = -\frac{qR + \cos kR}{kR + \sin kR}, \text{ for even } l,$$

$$\cot\eta_l = -\frac{qR - \cos kR}{kR - \sin kR}, \text{ for odd } l. \qquad (36)$$

It is evident that the total cross section of scattering by two centers calculated with these phases (because of their independence on $l$), as in the case (20), is equal to infinity. Indeed, rewriting the sum of partial cross sections as two infinite sums, we obtain

$$\sigma(k) = \frac{4\pi}{k^2}\sum_{l=0}^{\infty}(2l+1)\sin^2\eta_l = \frac{4\pi}{k^2}\left[\sum_{l=even}^{\infty}(2l+1)\sin^2\eta_l + \sum_{l=odd}^{\infty}(2l+1)\sin^2\eta_l\right] = \infty. \qquad (37)$$

While the cross section (32) obtained from the optical theorem is finite. The reason for this contradiction is the same as in Sec. 2, namely: refusal from the Huygens-Fresnel picture of scattering, use of the SSW asymptotic of the wave function and straightforward application of the usual S-matrix method for spherically symmetrical potential to targets having no spherical symmetry.

**4. Method of partial waves for non-spherical targets**

In the connection with the above stated a question arises: Is it possible to adapt the method of partial waves for the case of a multicenter target keeping the Huygens-Fresnel picture of the scattering process according to which far from the target there is a system of the secondary waves diverging from each of the centers? Demkov and Rudakov gave the positive answer to this question in their paper [23] where it was shown that the S-matrix method could be also applied to non-spherical potentials.

Let us briefly describe the main ideas of this method. The wave function describing elastic scattering of a particle by a spherically symmetrical potential is defined by the expression (1) where the radial part of the wave function has the asymptotic form (2). A molecular potential as a cluster of non-overlapping spherical potentials centered at the atomic sites is a non-spherical potential. In the Schrödinger equation with this potential it is impossible to separate the angular variables and represent the wave function at an arbitrary point of space in the form of expansion in spherical functions (1). However, asymptotically at great distances from the molecule the wave function can be written as expansion in a set of other orthonormal functions $Z_\lambda(\mathbf{k})$:

$$\psi_{\mathbf{k}}^+(\mathbf{r}\rightarrow\infty) \approx 4\pi\sum_\lambda R_{k\lambda}(r) Z_\lambda^*(\mathbf{k}) Z_\lambda(\mathbf{r}) \qquad (38)$$

with the radial part of the wave function

$$R_{k\lambda}(r \to \infty) \approx e^{i(\eta_\lambda + \frac{\pi}{2}\omega_\lambda)} \frac{1}{kr} \sin(kr - \frac{\pi}{2}\omega_\lambda + \eta_\lambda) \,. \tag{39}$$

Here the index $\lambda$ numerates different partial functions similar to the quantum numbers $l$ and $m$ for the central field; $\omega_\lambda$ is the quantum number (parity) that is equal to the orbital moment $l$ for the spherical symmetrical case; $\eta_\lambda(k)$ are the molecular phases. The explicit form of functions $Z_\lambda(\mathbf{k})$, naturally, depends on a specific type of the target field, particularly on the number of atoms forming the target and on mutual disposition of the scattering centers in space, *etc*. The functions $Z_\lambda(\mathbf{k})$, like the spherical functions $Y_{lm}(\mathbf{k})$, create an orthonormal system and for this reason:

$$\int Z_\lambda^*(\mathbf{k}) Z_\mu(\mathbf{k}) d\Omega_k = \delta_{\lambda\mu} \,. \tag{40}$$

The scattering amplitude for a non-spherical target, according to [23], is given by the following expression

$$F(\mathbf{k},\mathbf{k'}) = \frac{2\pi}{ik} \sum_\lambda (e^{2i\eta_\lambda} - 1) Z_\lambda^*(\mathbf{k}) Z_\lambda(\mathbf{k'}) \,. \tag{41}$$

The total elastic scattering cross section, i.e. the cross section integrated over all directions of momentum of the scattered electron $\mathbf{k'}$, is defined by the formula

$$\sigma(\mathbf{k}) = \frac{(4\pi)^2}{k^2} \sum_\lambda |Z_\lambda(\mathbf{k})|^2 \sin^2 \eta_\lambda \,. \tag{42}$$

Of course, this cross section depends on mutual orientation of incident electron momentum $\mathbf{k}$ and molecule axes. The cross section averaged over all the directions of momentum of incident electron $\mathbf{k}$ is connected with the molecular phases $\eta_\lambda(k)$ by the following formula

$$\overline{\sigma}(k) = \frac{4\pi}{k^2} \sum_\lambda \sin^2 \eta_\lambda \,. \tag{43}$$

In the case of a spherical symmetrical target the formula (43) exactly coincides with the known formula for the total cross section of scattering. Indeed, in the case of the central field the index $\lambda$ is replaced by the quantum numbers $l$ and $m$. But the phase of scattering by the central field is independent of the magnetic number and therefore for the given value of the orbital moment $l$ it is necessary to summarize over all $m$. This results in the factor $(2l+1)$ under the summation sign in formula (43). The partial wave (39) and molecular phases $\eta_\lambda(k)$ are classified, according to [23], by their behavior for low electron energies, i.e. for $k \to 0$. In this limit the particle wavelength is great as compared with the target size and the function $Z_\lambda(\mathbf{k})$ tends to some spherical function $Y_{lm}(\mathbf{k})$. The corresponding phase is characterized in this limit by the following asymptotic behavior: $\eta_\lambda(k) \to k^{2l+1}$.

Let us apply the method of partial waves for non-spherical targets [23] to the scattering of a slow particle by two short-rang potentials. Following to Ref. [28] the scattering amplitude (28) we rewrite in the form

$$F(\mathbf{k},\mathbf{k'},\mathbf{R}) = -\frac{2}{a+b}\cos(\mathbf{k}\cdot\mathbf{R}/2)\cos(\mathbf{k'}\cdot\mathbf{R}/2) + \frac{2}{a-b}\sin(\mathbf{k}\cdot\mathbf{R}/2)\sin(\mathbf{k'}\cdot\mathbf{R}/2) \,. \tag{44}$$

According to [23,28], the amplitude (44) should be considered as a sum of two partial amplitudes. The first of them is written as

$$\frac{4\pi}{2ik}(e^{2i\eta_0} - 1) Z_0(\mathbf{k}) Z_0^*(\mathbf{k'}) = -\frac{2}{a+b}\cos(\mathbf{k}\cdot\mathbf{R}/2)\cos(\mathbf{k'}\cdot\mathbf{R}/2) \,. \tag{45}$$

The second is defined by the following expression

$$\frac{4\pi}{2ik}(e^{2i\eta_1} - 1) Z_1(\mathbf{k}) Z_1^*(\mathbf{k'}) = \frac{2}{a-b}\sin(\mathbf{k}\cdot\mathbf{R}/2)\sin(\mathbf{k'}\cdot\mathbf{R}/2) \,. \tag{46}$$

The reasons for which we assigned the indexes at the functions $Z_\lambda(\mathbf{k})$ the values $\lambda = 0, 1$ will become understandable further. From formulas (45) and (46) after elementary transformations we obtain *two* molecular phases of scattering

$$\cot\eta_0 = \frac{\text{Re}[(a+b)^*]}{\text{Im}[(a+b)^*]} = -\frac{qR + \cos kR}{kR + \sin kR}, \quad \cot\eta_1 = \frac{\text{Re}[(a-b)^*]}{\text{Im}[(a-b)^*]} = -\frac{qR - \cos kR}{kR - \sin kR}. \tag{47}$$

These phases coincide with (36) but the principle difference is that the number of phases (36) is equal to infinity while in (47) there are two phases only. Substituting the phase shifts (47) in formulas (45) and (46), we obtain the functions $Z_\lambda(\mathbf{k})$ in the explicit form. They are defined by the following expressions

$$Z_0(\mathbf{k}) = \frac{\cos(\mathbf{k}\cdot\mathbf{R}/2)}{\sqrt{2\pi S_+}}, \qquad Z_1(\mathbf{k}) = \frac{\sin(\mathbf{k}\cdot\mathbf{R}/2)}{\sqrt{2\pi S_-}}. \tag{48}$$

Here $S_\pm = 1 \pm j_0(kR)$. It is easy to make sure that the functions (48) obey the conditions (40). It is evident that the functions (48) are defined by a geometrical target structure, i.e. by the direction of the molecular axis $\mathbf{R}$ in the arbitrary coordinate system in which the electron momentum vectors before and after scattering are $\mathbf{k}$ and $\mathbf{k}'$, respectively.

Study now the asymptotical behavior of the wave function (38) and (39). For this we write the exponent $\exp(i\mathbf{k}\cdot\mathbf{r})$ in the formula (26) as expansion in functions $Z_\lambda(\mathbf{k})$:

$$e^{i\mathbf{k}\cdot\mathbf{r}} = \sum_\lambda c_\lambda Z_\lambda(\mathbf{k}). \tag{49}$$

Multiplying the both parts of this equality by $Z_\mu^*(\mathbf{k})$ and integrating over all angles of the vector $\mathbf{k}$, we obtain the following expressions for the coefficients of the expansion (49):

$$c_0 = \sqrt{\frac{2\pi}{S_+}}[j_0(k|\mathbf{r}+\mathbf{R}/2|) + j_0(k|\mathbf{r}-\mathbf{R}/2|)],$$

$$c_1 = -i\sqrt{\frac{2\pi}{S_-}}[j_0(k|\mathbf{r}+\mathbf{R}/2|) - j_0(k|\mathbf{r}-\mathbf{R}/2|)]. \tag{50}$$

For great distances from the target the expansion coefficients in equations (50) have the form

$$c_0(r\to\infty) \approx 4\pi\frac{\sin kr}{kr}Z_0^*(\mathbf{k}) \quad \text{and} \quad c_1(r\to\infty) \approx -i4\pi\frac{\cos kr}{kr}Z_1^*(\mathbf{k}). \tag{51}$$

Consider now the asymptotical behavior of the partial wave with the index $\lambda = 0$. Taking into account the formulas (26), (45), (49) and (51), we write the corresponding partial wave from formula (38) in the form

$$4\pi R_{k0}(r)Z_0(\mathbf{k})Z_0^*(\mathbf{k}) = \frac{4\pi}{kr}[\sin kr + \frac{1}{2i}(e^{2i\eta_0}-1)e^{ikr}]Z_0(\mathbf{k})Z_0^*(\mathbf{k}). \tag{52}$$

From equation (52) immediately we obtain

$$R_{k0}(r\to\infty) = e^{i\eta_0}\frac{1}{kr}\sin(kr+\eta_0). \tag{53}$$

Making the same operations for the case $\lambda = 1$, we obtain for the second partial wave the following expression

$$R_{k1}(r\to\infty) = e^{i(\eta_1+\frac{\pi}{2})}\frac{1}{kr}\sin(kr-\frac{\pi}{2}+\eta_1). \tag{54}$$

Comparing these expressions with the general formula (39) we come to the following conclusions. If the electron states are characterized by a projection of the angular momentum on the $\mathbf{R}$ axis and by parity of the wave function relative to the reflection in the plane perpendicular to $\mathbf{R}$ and going through the middle of the inter-atomic distance, then the first of the partial waves (53) corresponds to the state $\Sigma_g$ and the second one (54) to $\Sigma_u$. The molecular phases $\eta_\lambda(k)$ can be classified by considering their behavior for $k\to 0$ [23]. In this limit the electron wavelength is much greater than the target size and the picture of scattering has to approach to the spherical symmetry one. Consider this limit transition in the formulas (47); we obtain: $\eta_0(k\to 0)\sim k$ and $\eta_1(k\to 0)\sim k^3$. Thus, the molecular phases behave similar to the *s* and *p* phases in the spherically symmetrical potential. By this is explained the choice of their indexes.

The transition to the limit $k \to 0$ in formulas (48) gives instead of the functions $Z_\lambda(\mathbf{k})$ the well-known spherical functions

$$Z_0(\mathbf{k})_{k \to 0} \to \frac{1}{\sqrt{4\pi}} \equiv Y_{00}(\mathbf{k}), \ Z_1(\mathbf{k})_{k \to 0} \to \sqrt{\frac{3}{4\pi}} \cos\vartheta \equiv Y_{10}(\mathbf{k}). \qquad (55)$$

Here $\vartheta$ is the angle between the vector **k** and axis **R**.

Finally, substituting the molecular phases (47) in the formula (43) for partial cross section [28]

$$\bar{\sigma}(k) = \frac{4\pi}{k^2}[\sin^2\eta_0 + \sin^2\eta_1] = \frac{4\pi}{k^2}[(1 + \cot^2\eta_0)^{-1} + (1 + \cot^2\eta_1)^{-1}]. \qquad (56)$$

we have the cross section of elastic scattering (32) that was obtained early with the help of the optical theorem.

Summarize the results obtained with the method of partial waves for a target formed by two non-overlapping atomic potentials. For such targets the molecular phases of scattering and the functions $Z_\lambda(\mathbf{k})$ can be found explicitly. The phases of molecular scattering $\eta_\lambda(k)$, as the atomic ones $\delta_l(k)$, are the functions of electron momentum $k = |\mathbf{k}|$ only, as in the usual S-matrix method. The form of the functions $Z_\lambda(\mathbf{k})$ is defined by the structure of a target and its orientation in space. The number of non-zero molecular phases in this case is equal to two. This is connected with the fact that each of these two scattering centers is a source of *s*-spherical waves only, which is valid for the case of low electron energy. If the scattering by each of these centers would be accompanied by generation of spherical waves with non-zero orbital moments $l = 0 \div l_{max}$ then the number of non-zero molecular phases $\eta_\lambda(k)$ is equal to $N(l_{max} + 1)^2$. Here *N* is the number of scatterers [29].

It is simple to calculate the functions $Z_\lambda(\mathbf{k})$ and the scattering phases $\eta_\lambda(k)$ when one knows the exact wave function (26). On the other hand, if this function is known, as it is the case for a system of non-overlapping potentials, the scattering amplitude can be obtained in the closed form [29] and the cross section can be found with the help of the optical theorem, and therefore there is necessity to resort to the method of partial waves. However, for non-spherical potentials different from muffin-tin-potentials, the application of the partial wave method [23] makes it possible to separate in the explicit form the scattering dynamics contained in the molecular phases $\eta_\lambda(k)$ from the kinematics of the process and from target structure defined by the functions $Z_\lambda(\mathbf{k})$.

## 5. Conclusions

It has been demonstrated that the assumption on SSW asymptotic of the wave function of the molecular continuum, underlying the MS and SCE methods, leads to unavoidable contradictions and therefore cannot be considered as correct. Refusal from the Huygens-Fresnel picture of scattering is equivalent, *per se*, to replacement of diffraction by a system of non-overlapping potentials with electron wave diffraction by a *molecular sphere* (in the MS picture) or by a *system of concentric molecular spheres* (in the SCE picture). It has been shown that the method of partial waves for non-spherical targets ought to be constructed according to the general theory [23] rather than by means of straightforward application of the usual S-matrix method developed for spherically symmetrical potentials.

The use of non-overlapping atomic potentials or muffin-tin-potentials for molecule description assumes that the continuum wave functions inside atomic spheres are known. Hence, one knows the sets of the scattering phases $\delta_l(k)$ for each of the atomic potentials. In this case, as shown in [29], it is possible to obtain the elastic scattering amplitude in closed form and the calculation of this amplitude reduces to solving a system of non-homogeneous algebraic equations. Therefore, within the model of the non-overlapping atomic potentials there is no necessity to resort to the method of partial waves.

**Acknowledgements**
This work was supported by DAAD (Deutscher Akademischer Austausch Dienst) and Uzbek Foundation Award ФА-Ф2-Ф100.

**Figure captures**
Fig. 1. *a* – the MS and SCE scattering picture: *b* – the scattering picture according to the Huygens-Fresnel principle.

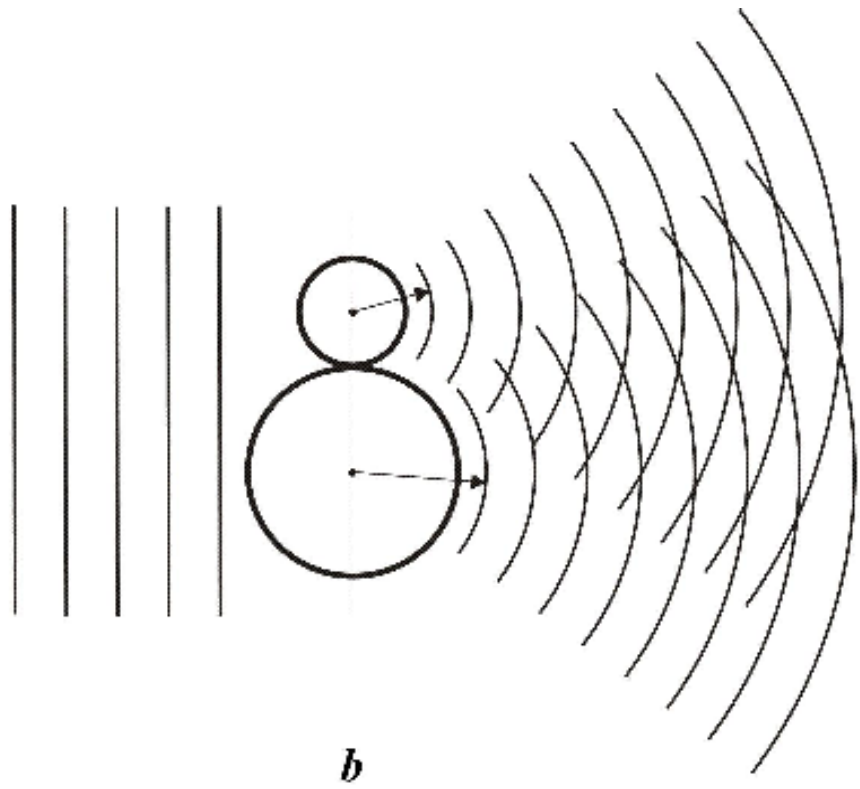

*b*

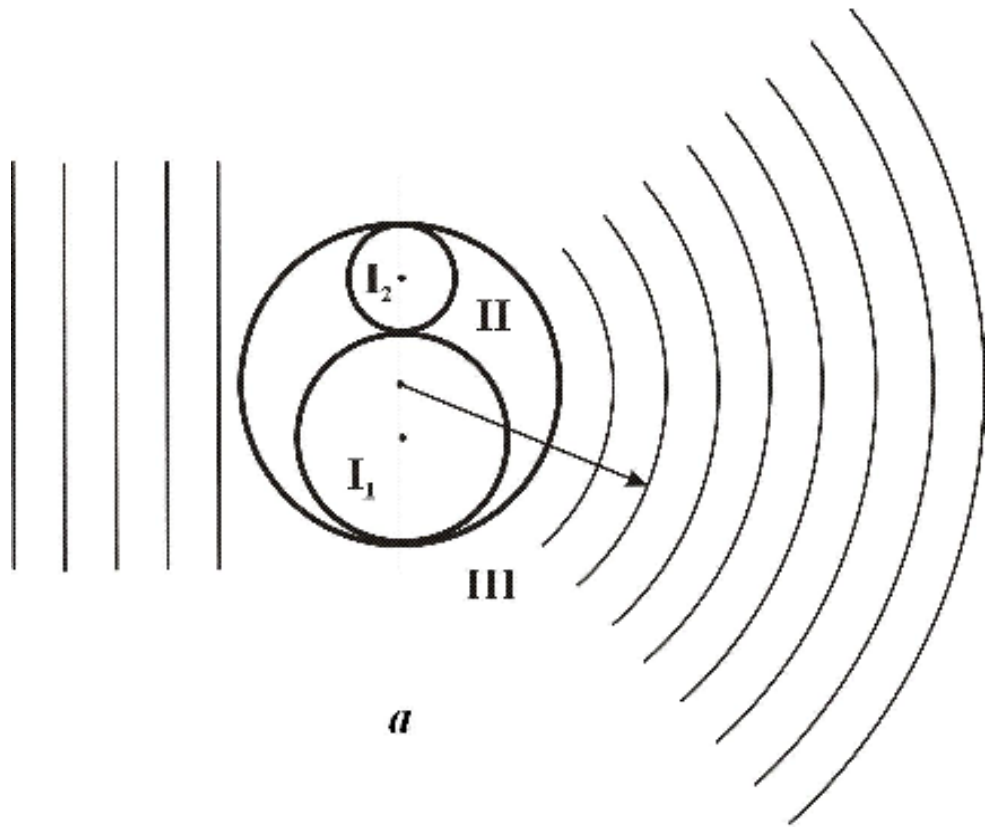

*a*